\documentstyle[emulateapj,apjfonts,psfig]{article}

\lefthead{J.~M.~Miller et al.}  \righthead{XTE~J1650$-$500}

\begin{document}

\title{Evidence for Spin and Energy Extraction in a Galactic Black
Hole Candidate:\\ The \textit{XMM-Newton}/EPIC-pn Spectrum of 
XTE~J1650$-$500}

\author{J.~M.~Miller\altaffilmark{1}, 
        A.~C.~Fabian\altaffilmark{2},
	R.~Wijnands\altaffilmark{1,3},
	C.~S.~Reynolds\altaffilmark{4},
	M.~Ehle\altaffilmark{5},
	M.~J.~Freyberg\altaffilmark{6},\\
	M.~van~der~Klis\altaffilmark{7},
	W.~H.~G.~Lewin\altaffilmark{1},
	C.~Sanchez-Fernandez\altaffilmark{8},
	A.~J.~Castro-Tirado\altaffilmark{9}
	}

\altaffiltext{1}{Center~for~Space~Research and Department~of~Physics,
        Massachusetts~Institute~of~Technology, 70 Vassar Street,
        Cambridge, MA 02139--4307; jmm@space.mit.edu}
\altaffiltext{2}{Institute of Astronomy, University of Cambridge,
        Madingley Road, Cambridge CB3 OHA, UK}
\altaffiltext{3}{\it Chandra Fellow}
\altaffiltext{4}{Department of Astronomy, University of Maryland,
        College Park, MD, 20742}
\altaffiltext{5}{\textit{XMM-Newton} SOC, Villafranca Satellite
        Tracking Station, PO Box 50727, 28080, Madrid, ES \& Research
        and Scientific Support Dept. of ESA, Noordwijk, NL}
\altaffiltext{6}{Max-Planck-Institut f\"ur extraterrestrische Physik,
        Postfach 1312, 85741 Garching, DE}
\altaffiltext{7}{Astronomical Institute ``Anton Pannekoek,''
        University of Amsterdam, and Center for High Energy
        Astrophysics, Kruislaan 403, 1098 SJ, Amsterdam, NL}
\altaffiltext{8}{Laboratorio de Astrof\'{i}sica Espacial y F\'{i}sica
        Fundamental (LAEFF-INTA), PO Box 50727, E-28080, Madrid, ES}
\altaffiltext{9}{Instituto de Astrof\'{i}sica de Andaluc\'{i}a
        (IAA--CSIC), P. O. Box 03004, E18080--Granada, ES \&
        LAEFF--INTA, Madrid, ES}

\authoremail{jmm@space.mit.edu}

\label{firstpage}

\begin{abstract}
We observed the Galactic black hole candidate XTE~J1650$-$500 early in
its Fall, 2001 outburst with the \textit{XMM-Newton} European Photon
Imaging pn Camera (EPIC-pn).  The observed spectrum is consistent with
the source having been in the ``very high'' state.  We find a broad,
skewed Fe~K$\alpha$ emission line which suggests that the primary in
this system may be a Kerr black hole, and which indicates a steep disk
emissivity profile that is hard to explain in terms of a standard
accretion disk model.  These results are quantitatively and
qualitatively similar to those from an \textit{XMM-Newton}
observation of the Seyfert galaxy MCG--6-30-15.  The steep emissivity
in MCG--6-30-15 may be explained by the extraction and dissipation of
rotational energy from a black hole with nearly-maximal angular
momentum or material in the plunging region via magnetic connections
to the inner accretion disk.  If this process is at work in both
sources, an exotic but fundamental general relativistic prediction may
be confirmed across a factor of $10^{6}$ in black hole mass.  We
discuss these results in terms of the accretion flow geometry in
stellar-mass black holes, and the variety of enigmatic phenomena often
observed in the very high state.
\end{abstract}

       
 \section{Introduction}
The accretion flow geometry in stellar-mass black holes may change
considerably with the mass accretion rate ($\dot{m}$).  Characteristic
periods of correlated intensity, energy spectral hardness, and fast
variability in the X-ray band are identified as ``states'' (for a
review, see Tanaka \& Lewin 1995; see also Homan et al. 2001).  These
states are observed in all stellar-mass black holes, and are thought
to be driven primarily by changes in $\dot{m}$.

Of five common states, the ``very high'' state is perhaps the
least-understood.  Often the first state observed in a transient
outburst and usually very luminous, a variety of phenomena are
observed that indicate a unique inner accretion flow environment.
Fast quasi-periodic oscillations (QPOs; 30--450~Hz) --- if tied to the
Keplerian frequency of the inner accretion disk --- indicate accretion
disks which extend close to the marginally stable circular orbit for a
Schwarzschild black hole, and may even indicate black hole spin
(e. g., Strohmayer 2001, Miller et al. 2001a).  Discrete jet ejections
are sometimes observed in the radio band with velocities that approach
$c$ (for a review, see Fender 2001).  The X-ray energy spectra
observed in the very high state are a mix of thermal and non-thermal
components.  The spectra suggest a disk which may extend close to the
black hole, and a corona (the assumed non-thermal source) which
irradiates the disk and produces a ``reflection'' spectrum (George \&
Fabian 1991; see also Gierlinski et al. 1999).

With adequate spectral resolution, Fe~K$\alpha$ line profiles may
provide effective tools for constraining the nature of accretion flows
(e.\ g.\ Miller et al. 2002) in stellar-mass black holes.  Such
profiles have been used to infer black hole spin and rotational energy
extraction via magnetic fields in active galactic nuclei (AGNs;
Iwasawa et al. 1996, Wilms et al. 2001).  Evidence for spin, based on
Fe~K$\alpha$ line profiles, has also been reported in stellar-mass
black holes, but with gas proportional counters offering lower
resolution and inhibiting stronger conclusions (Balucinska-Church \&
Church 2000, Miller et al. 2001b, Campana et al. 2002).  The high
effective area of \textit{XMM-Newton} and short frame times available
with the EPIC-pn camera are well-suited to capturing high resolution
spectra of bright sources.

We were granted a target-of-opportunity observation to study the
Galactic black hole candidate XTE~J1650$-$500 (Remillard 2001) early
in its Fall, 2001 outburst.  Although a neutron star primary can only
be ruled-out by optical radial velocity measurements, there is strong
evidence in X-rays that XTE~J1650$-$500 harbors a black hole.  Based
on the spectrum we obtained with \textit{XMM-Newton} and timing
properties (Wijnands, Miller, \& Lewin 2001), we observed
XTE~J1650$-$500 in the very high state --- the first pile-up-free CCD
spectrum of a stellar-mass black hole in this state.  Herein, we
report the detection of a broad Fe~K$\alpha$ line.  We discuss the
implications of this line for the accretion flow geometry in this
source, and explore connections to the recent \textit{XMM-Newton}
observation of the Seyfert-I AGN MCG--6-30-15 (Wilms et al. 2001).

\section{Observation and Data Reduction}
We observed XTE~J1650$-$500 with \textit{XMM-Newton} on 13 September,
2001, from 15:45:25--21:41:04 (UT) for a total exposure of 21.4~ks.
We did not operate the EPIC-MOS2 camera or the Optical Monitor to
preserve telemetry.  The EPIC-MOS1 camera was operated in timing mode
but suffered a full scientific buffer nearly continuously due to the
high flux.  Full spectral results from the Reflection Grating
Spectrometer (RGS, 0.33--2.5~keV) will be reported in a separate
paper.  To prevent photon pile-up, the EPIC-pn camera was operated in
``burst'' mode with the ``thin'' optical filter in place (for more
information on the pn camera, see St\"uder et al. 2001).  Only one CCD
of the pn camera is active in burst mode, and spatial information is
only recorded in one dimension.  Burst mode has a time resolution of
7~$\mu$s, but a duty cycle of approximately 3\%.

We extracted source counts using a box region centered on the source
position (with X an Y half-widths of 1694 and 28, respectively, in DET
units).  Two background regions adjacent to the source position were
selected.  Single and double events were included for analysis.
Standard pn filtering was accomplished with the procedure XMMEA\_EP,
assuming a source position of $16^{\circ} 50' 01.0'', -49^{h} 57^{m}
45^{s}$ (Castro-Tirado et al. 2001; Groot et al. 2001).  We measured a
source count rate of $2557.0\pm 1.9$~counts~s$^{-1}$, and a background
count rate of $151.5\pm 0.5$~counts~s$^{-1}$.  Calibrated event lists
were custom-made at MPE using SAS--5.3$\alpha$.  The source and
background spectra were created with SAS--5.2.0.  Spectra were grouped
based both on a requirement for counts per channel (20) and on the
maximum number of spectral channels sampling the pn energy resolution
(3).

This is the first spectral analysis to be reported from an observation
which employed burst mode.  Despite the peculiarities of this mode,
the instrumental effective area and the energy response, and charge
transfer inefficiency (CTI) correction do not differ from more
standard modes (Kirsch et al. 2002).  Therefore, ``full frame'' mode
response matrices were created to model the instrument response using
SAS--5.2.0.  

The spectral fits reported herein were made using XSPEC version 11.1.0
(Arnaud 1996).  Quoted errors correspond to $\Delta~\chi^{2}=1.0$.
Below 0.5~keV and above 10.0~keV, deviations were seen in the
residuals regardless of the spectral model used.  We therefore
restricted our analysis to the 0.5--10.0~keV band.  We note a feature
at 2.34~keV which appears as an emission line.  The detector effective
area changes sharply at this energy and the line is well-fit by a
zero-width Gaussian, suggesting that the feature is instrumental.

\section{Analysis and Results}
In examining the EPIC-pn and RGS data of XTE~J1650$-$500, we find
oxygen to be $13^{+1}_{-5}$\% under-abundant, neon to be
$16_{-6}^{+8}$\% over-abundant, and iron to be $45\pm 5$\%
under-abundant relative to solar values along this line of sight
(assuming abundances as measured by Anders \& Grevesse 1989).  The RGS
spectrum reveals that the oxygen edge location is more consistent with
0.536~keV than the expected 0.532~keV; this has also been noted in
fits to Chandra data of Cygnus~X-1 (Schulz et al. 2001, Miller et
al. 2002).  We therefore fixed the oxygen edge to be at 0.536~keV in
all fits.  All other elements are found to have abundances consistent
with solar values.

Absorption in the ISM was modeled using the ``vphabs'' model in XSPEC
(with the abundance of oxygen set to zero,

\centerline{~\psfig{file=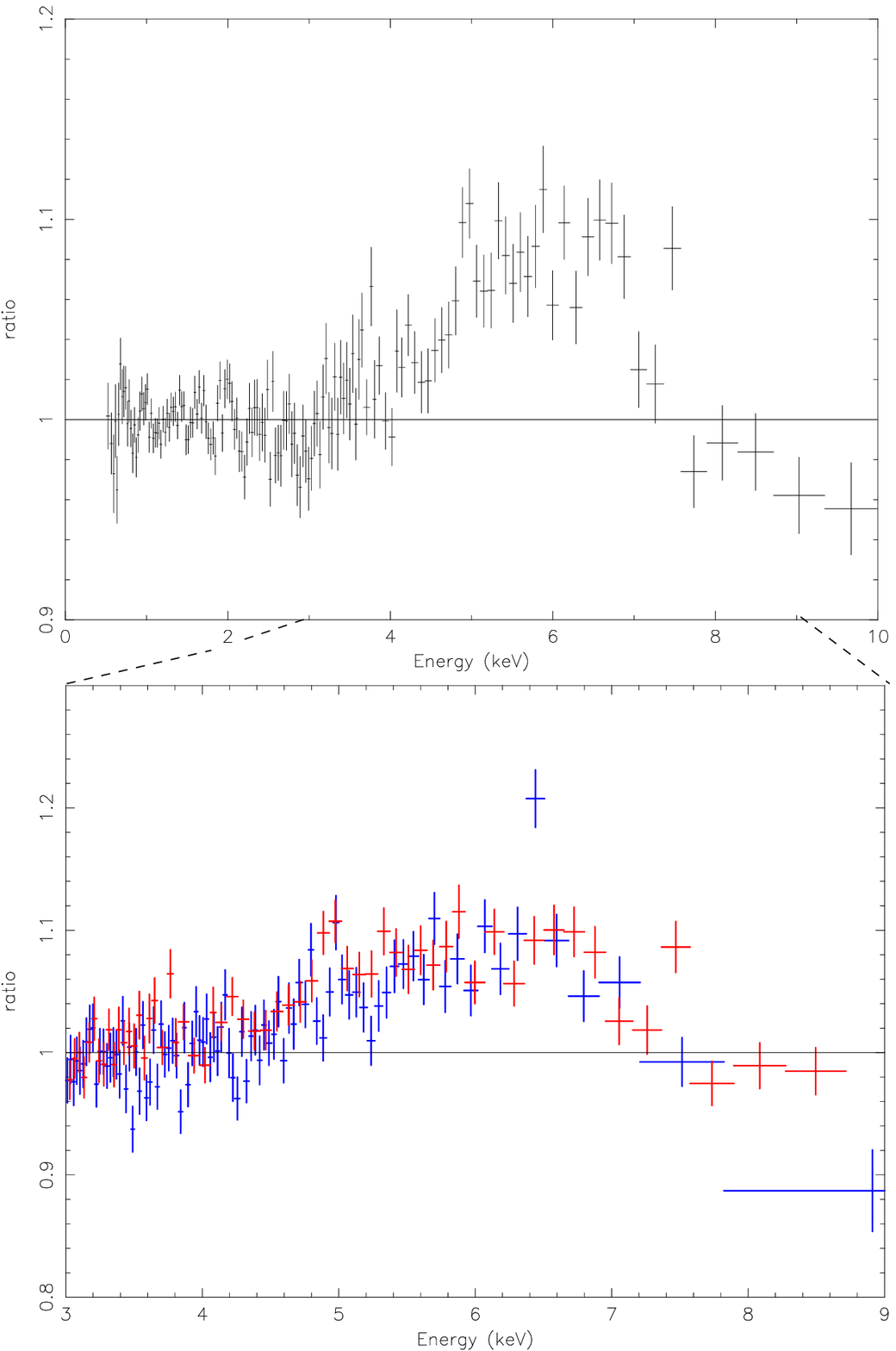,height=4.0in}~}
\figcaption[]{\scriptsize Above: The data/model ratio for a model
consisting of multicolor disk blackbody and power-law components,
modified by photoelectric absorption.  (We have suppressed a feature
at 2.34~keV; the data is fit between 0.5--10.0~keV) Below: The
data/model ratio of XTE~J1650$-$500 (in red) and that for a 30~ksec
observation of Cygnus X-1 with the \textit{Chandra} High Energy
Transmission Grating Spectrometer (in blue), shown in greater detail
than above.  In all ratios shown above, the 4.0--7.0~keV band was
ignored in fitting the model.  The ratios have been rebinned for
visual clarity.  In both, a majority of the line profile lies below
6.40~keV, the K$\alpha$ line energy for neutral Fe.}
\medskip

and an additional edge to model the oxygen absorption at 0.536~keV).
We measure an effective neutral hydrogen column density of
$N_{H}=7.8\pm 0.2\times 10^{21}$~atoms~cm$^{-2}$.

Fits with the standard multicolor accretion disk black body
(hereafter, MCD; Mitusda et al. 1984) plus power-law model to the
0.5--10.0~keV spectrum were statistically unacceptable
($\chi^{2}=513.4$ for 233 d.o.f.).  The data/model ratio for this
model is shown in Figure 1.  The emission line profile in this
data/model ratio is similar to that observed in Cygnus X-1 with the
\textit{Chandra} High Energy Transmission Grating Spectrometer (Miller
et al. 2002), and similar to line profiles seen in some AGNs with
\textit{ASCA} (see, e. g., Iwasawa et al. 1996; Weaver, Gelbord, \&
Yaqoob 2001).

Gaussian emission line and smeared edge (see Ebisawa et al. 1994)
components were added to the model.  The addition of these components
improved the fit significantly ($\chi^{2}=314.8$ for 225 d.o.f.).
Using this model, we measure an inner disk color temperature of
$kT=0.322\pm 0.004$~keV, and an MCD normalization of
$3.9^{+0.3}_{-1.3}\times 10^{4}$.  The measured power-law index is
$\Gamma=2.09^{+0.03}_{-0.09}$; the normalization of this component is
measured to be $3.1\pm 0.2$ ph~cm$^{-2}$~s$^{-1}$~keV$^{-1}$ at 1~keV.
The 0.5--10.0~keV flux of the MCD component is
$0.36^{+0.03}_{-0.12}\times 10^{-8}$~erg~cm$^{-2}$~s$^{-1}$, and that
for the power-law component is $1.17\pm 0.08\times
10^{-8}$~erg~cm$^{-2}$~s$^{-1}$.

The best-fit Gaussian indicates a broad line, shifted from the neutral
Fe~K$\alpha$ line energy of 6.40~keV: E$=5.3^{+0.1}_{-0.3}$~keV,
FWHM$=3.2_{-0.6}^{+0.8}$~keV, and W$=250\pm 50$~eV.  The edge is
measured to be at E$=6.8\pm 0.3$~keV, with a depth of $\tau=0.5\pm
0.1$.  However, significant residuals remain in the Fe~K$\alpha$ line
region with this fit, due in part to the non-Gaussian nature of the
line profile (see Figure 1).  Moreover, on a broader energy range (one
that includes the 20--30~keV ``Compton hump'' seen in many sources), a
reflection model is often required to fit the spectra of stellar-mass
black holes (see, e.g., Gierlinski et al. 1999).  Broad Gaussian and
smeared edge components are merely an approximation to a full
reflection model in the 0.5--10.0~keV band.  Therefore, we now focus
on the results of fitting more sophisticated, physically-motivated
reflection models.  These replace the hard power-law, Gaussian, and
smeared edge components discussed above.

Anticipating a highly-ionized accretion disk, we made fits with the
``constant density ionized disk'' reflection model (hereafter, CDID;
Ross, Fabian, \& Young 1999).  This model measures the relative
strengths of the directly-observed and reflected flux, the accretion
disk ionization parameter ($\xi = L_{X}/nR^{2}$, where $L_{X}$ is the
X-ray luminosity, $n$ is the hydrogen number density, and $R$ is
radius), and the photon index of the illuminating power-law flux.
Fe~K$\alpha$ line emission and line broadening due to Comptonization
in an ionized disk surface layer are included in this model.

The fit obtained with this model is shown in the top panel of
Figure~2.  The photon index of the irradiating power law is measured
to be $\Gamma=2.08^{+0.02}_{-0.04}$.  The ionization parameter is
high: $\xi=1.3^{+0.7}_{-0.1}\times 10^{4}$~erg~cm~s$^{-1}$, and the
relative strength of reflected flux is measured to be
$f=0.5_{-0.1}^{+0.7}$ (where $F_{total} = F_{direct} + f\times
F_{refl.}$).  While the shape of the Fe~K edge is reproduced by this
model, the width and shape of the Fe~K$\alpha$ line is not, and a
statistically-poor fit is obtained ($\chi^{2}=399.7$ for 231 d.o.f.).
The ionization parameter obtained with this model corresponds to a
mixture of helium-like and hydrogenic ion species of Fe (Kallman and
McCray 1982).

As Doppler shifts and general relativistic smearing may be expected
for lines produced in an accretion disk close to the black hole, we
next made fits after convolving (or, ``blurring'') the CDID model with
the line element expected near a Kerr black hole.  We assumed an inner
disk radius of 1.24~$R_{g}$, an outer line production radius of
400~$R_{g}$, and an inclination of $i=45^{\circ}$ (in fits with this
and other models, intermediate inclinations were marginally preferred
in terms of $\chi^{2}$).  With this blurred model, we obtained
parameter constraints which differed marginally from the previous fit:
$\xi=2.5^{+5.5}_{-0.1}\times 10^{4}$~erg~cm~s$^{-1}$,
$f=0.6^{+0.6}_{-0.1}$, and $\Gamma=1.96^{+0.04}_{-0.06}$.  The shape
and strength of the Fe~K$\alpha$ line are not fit adequately; the fit
is slightly worse than the un-blurred model ($\chi^{2}=407.8$, 231
d.o.f.).

Finally, we constructed a model which allows the Fe~K$\alpha$ line and
reflection components to be treated separately.  We fit the line with
the ``Laor'' line model (Laor 1991), and the power-law and reflection
continuum (minus the line) with the ``pexriv'' model (Magdziarz \&
Zdziarski 1995).  With pexriv, $f=1$ corresponds to a disk which
intercepts half of the incident power-law flux.  It should be noted
that this model was also used by Wilms et al. (2001) in fits to the
\textit{XMM-Newton}/EPIC spectrum of the Seyfert galaxy MCG--6-30-15,
allowing for a direct comparison.  For the Laor line, we initially
fixed the inner disk edge at 1.24~$R_{g}$, the outer line production
region at 400~$R_{g}$, and the inclination at $i=45^{\circ}$.  The
line energy, emissivity profile ($\epsilon \sim r^{-\beta}$; we fit
for $\beta$), and intensity were allowed to vary.  The MCD and pexriv
reflection components were blurred as before.

We found that the data could not simultaneously constrain $f$, $\xi$,
and the disk surface temperature with pexriv (an additional

\centerline{~\psfig{file=f2.epsi,height=2.0in,angle=-90}~}
\centerline{~\psfig{file=f3.epsi,height=2.0in,angle=-90}~}
\figcaption[]{\scriptsize \textit{Top}: the spectrum of
XTE~J1650$-$500 fit with a model for disk reflection (shown in red is
the ``constant density ionized disk'' model; Ross, Fabian, \& Young
1999).  Clearly, the line profile is broader than is predicted with
this model, indicating significant Doppler shifts and/or general
relativistic skewing are required to describe the line profile.
\textit{Bottom}: The spectrum fit with the ``Laor'' model for line
emission near a Kerr black hole (Laor 1991), and the ``pexriv''
reflection model (Magdziarz \& Zdziarski 1995).}
\medskip
 
parameter for this model).  We therefore fixed the ionization
parameter at $\xi=2.0\times 10^{4}$~erg~cm~s$^{-1}$, and the disk
surface temperature at $kT=1.3~$keV~(as per Ross, Fabian, \& Young
1999 in fits to Cygnus X-1, wherein a similarly low MCD disk
temperature but similarly high values of $\xi$ are reported).  

The fit obtained with this model is shown in the bottom panel of
Figure 2.  Statistically, this model represents a significant
improvement ($\chi^{2}=319.9$ for 229 d.o.f.).  A power-law index of
$\Gamma=2.04^{+0.03}_{-0.02}$ is obtained.  We measure the
Fe~K$\alpha$ line to be centered at E$=6.8^{+0.2}_{-0.1}$~keV, likely
due to a blend of Fe~XXV and Fe~XXVI (helium-like and hydrogenic Fe)
and consistent with the high ionization parameters previously
measured.  The line is strong, with an equivalent width of
W$=350^{+60}_{-40}$~eV and a flux of $2.2\pm 0.3\times
10^{-10}$~erg~cm$^{-2}$~s$^{-1}$~($3.2\pm 0.4\times
10^{-2}$~ph~cm$^{-2}$~s$^{-1}$).  When the inner disk edge is allowed
to vary, an inner radius of 1.24~$R_{g}$~(the limit of the Laor model,
corresponding to $a=0.998$) is preferred over an inner radius of
6.0~$R_{g}$~(the marginally stable circular orbit around a
Schwarzschild black hole) at the 6$\sigma$ level of confidence.  A
steep emissivity is suggested via the Laor line model: $\beta=5.4\pm
0.5$.  This emissivity is preferred over that for a standard accretion
disk ($\beta=3.0$) at the 5.6$\sigma$ level of confidence.  The pexriv
reflection ``fraction'' is $f=0.6_{-0.1}^{+0.3}$.  This is consistent
with the reflection fractions measured in the high state of Cygnus~X-1
(Gierlinski et al. 1999), wherein the disk may extend to the
marginally stable circular orbit.

It is not likely that these results can be explained in terms of an
anomalous Fe abundance.  Allowing $\beta$ and $A_{Fe}$ (in pexriv) to
vary, the abundance is poorly constrained: $A_{Fe} =
1.0^{+0.5}_{-0.7}$, and the emissivity drops only slightly to
$\beta=5.0$.  Fixing the emissivity at $\beta=3.0$ and allowing the Fe
abundance to vary yields a significantly worse fit ($\chi^{2}=353.1$
for 229 d.o.f.) and Fe must be more than 30 times over-abundant.  We
note that the line strength and reflection fraction in our final fit
are not strictly congruent, and that $f=0.6^{+0.3}_{-0.1}$ is below
the reflection fraction measured in MCG--6-30-15 ($f=1.5-2.0$, Wilms
et al.\ 2001) using a similar model.  If we fix the $f=1.5$ in our
final model, $\chi^{2}$ increases only slightly ($\chi^{2}=324.9$
for 230 d.o.f.), and the other fit parameters only change within their
1$\sigma$ confidence intervals.  Thus, more congruent values of $f$
are allowed by the data.

The high $\chi^{2}$ value associated with our final model could be due
to unmodeled narrow spectral features, approximations in the model,
and calibration issues.  We note a feature at approximately 7.0~keV,
which may be a narrow edge due to neutral Fe or an Fe~XXVI absorption
line.  Alternatively, there may be an Fe~K$\beta$ emission line near
7.4~keV due to ionized Fe species, which is superimposed upon the
broader smeared edge fit by the reflection models.  There is also weak
evidence for a narrow absorption edge feature near 9.3~keV, consistent
with Fe~XXVI.  Modeling these features and the addition of 0.5\%
systematic errors below 1~keV are sufficient to make the fit
acceptable.

\section{Discussion}
We have observed a broad Fe~K$\alpha$ line profile in the
\textit{XMM-Newton}/EPIC-pn spectrum of the Galactic black hole
candidate XTE~J1650$-$500 in the very high state.  A comparison with
the broad line profile observed with \textit{Chandra} in Cygnus X-1 is
shown in Figure~1.  That such profiles are observed in very different
systems, strongly suggests that broad Fe~K$\alpha$ lines in
stellar-mass black holes stem from a common process.  Like the broad
lines observed in some Seyfert AGNs, these lines are likely produced
by irradiation of the inner disk.

The Fe~K$\alpha$ line we have observed in XTE~J1650$-$500 suggests a
Kerr black hole with near-maximal angular momentum ($a=0.998$).  The
accretion disk emissivity profile measured with the Laor line model is
inconsistent with the energy dissipation expected for standard disks.
These results are very similar to those reported by Wilms et
al. (2001) using the same line and reflection models for the broad
Fe~K$\alpha$ line observed in an \textit{XMM-Newton}/EPIC-pn spectrum
of the Seyfert galaxy MCG--6-30-15 (E$=6.97_{-0.10}$~keV,
W$\sim300$--400~eV, $f=1.5-2$, $\beta\sim4.3$--5.0).  Those authors
suggest that rotational energy extraction from the spinning black hole
(Blandford \& Znajek 1977) or material in the plunging region (Agol \&
Krolik 2000) may infuse the inner accretion disk with extra energy via
magnetic connections, creating the steep emissivity profile indicated
by the Fe~K$\alpha$ line.  It is possible that rotational energy
extraction may be at work in XTE~J1650$-$500 as well.  If so, a
fundamental general relativistic prediction may be confirmed across a
factor of roughly $10^{6}$ in black hole mass.

This observation suggests a connection between the accretion geometry
of stellar-mass black holes in the very high state, and that inferred
in some Seyfert galaxies.  This is an important step towards
understanding the nature of the very high state, and the variety of
exotic phenomena observed in this state.  The Blandford--Znajek
process is also often invoked as a means of launching jets (Blandford
2001a, 2001b; see also Fender 2001).  That we have found an emissivity
which might be explained by magnetic connections to the black hole or
to matter in the plunging region in the very high state of
XTE~J1650$-$500 (detected at 7.5~mJy at 0.8~GHz with MOST in this
state with a spectrum indicative of jets; S. Tingay, priv. comm.),
suggests that the discrete radio ejections observed in some sources in
this state (see Fender 2001) may be driven by rotational energy
extraction.

Martocchia, Matt, \& Karas (2002) have shown that a ``lamp-post''
reflection model may explain the steep disk emissivity implied in fits
to the Fe~K$\alpha$ line in MCG--6-30-15.  This model assumes a source
of power-law flux which illuminates the accretion disk from a location
directly above the black hole.  To explain $\beta\sim4$, this model
requires $f\sim4$ --- well above the values we measure.  Therefore,
the lamp-post model may not adequately describe the accretion geometry
of XTE~J1650$-$500.

\section{Acknowledgments}
We wish to thank \textit{XMM-Newton} project scientist Fred Jansen for
executing our TOO request.  RW was supported by NASA through Chandra
fellowship grants PF9-10010, which is operated by the Smithsonian
Astrophysical Observatory for NASA under contract NAS8--39073.  This
work is based on observations obtained with \textit{XMM-Newton}, an
ESA science mission with instruments and contributions directly funded
by ESA Member States and the USA (NASA).

\end{document}